\documentclass[12pt]{article}
\begin{document}
%%%%%%%%%%%%%%%%%%%%%%%%%%%%%%%%%%%%%%%%%%%%%%%%%%%%%
%%%%%%%%%%%%%%%%%%%%%%%%%%%%%%%%%%%%%%%%%%%%%%%%%%%%%
%%%%%%%%%%%%%%%%%%%%%%%%%%%%%%%%%%%%%%%%%%%%%%%%%%%%%
%\baselineskip .25in
\begin{flushright}
OHSTPY-HEP-T-98-024 \\
%MPI H-V ?? \\
hep-th/9810158
\end{flushright}
\vspace{20mm}
\begin{center}
{\LARGE A Comment on the Light-Cone Vacuum in
1+1 Dimensional Super-Yang-Mills Theory}
\\
\vspace{20mm}
{\bf F.Antonuccio${}^{(a)}$
\footnote{anton@pacific.mps.ohio-state.edu},
  S.Pinsky${}^{(a)}$ 
\footnote{pinsky@mps.ohio-state.edu}, 
and
S.Tsujimaru${}^{(b)}$ 
\footnote{sho@tick.mpi-hd.mpg.de}} \\
\vspace{4mm}
{\em ${}^{(a)}$Department of Physics,\\ The Ohio State University,\\
Columbus,
OH 43210, USA\\
\vspace{4mm} and \\
\vspace{4mm}
${}^{(b)}$Max-Planck-Institut f\"{u}r Kernphysik, \\ 69029 Heidelberg,
Germany}

\end{center}
\vspace{10mm}

\begin{abstract}
The Discrete Light-Cone Quantization (DLCQ)
of a supersymmetric gauge theory in 1+1 dimensions
is discussed, with particular attention given to
the inclusion of the gauge zero mode. 
Interestingly, the notorious `zero-mode' problem
is now tractable because of special supersymmetric cancellations.
In particular, we show that
anomalous zero-mode contributions to the currents
are absent, in contrast to what is observed in
the non-supersymmetric case.
An analysis of the vacuum structure is provided
by deriving the effective quantum mechanical Hamiltonian
of the gauge zero mode. It is shown that the inclusion
of the zero modes of the adjoint scalars and fermions is  crucial
for probing the phase properties of the vacua. 
We find that the ground state energy is
zero and thus consistent with unbroken supersymmetry,
and conclude that the light-cone Fock vacuum is unchanged
with or without the presence of matter fields.

\end{abstract}
\newpage
%%%%%%%%%%%%%%%%%%%%%%%%%%%%%%%%%%%%%%%%%%%%%%
\section{Introduction}
A possibly surprising outcome of recent developments in string/M theory
are the proposed connections between non-perturbative
objects in string theory, and supersymmetric
gauge theories in low dimensions \cite{bfss97, dvv97}.
It is therefore of interest to study directly the non-perturbative
properties of super-Yang-Mills theories in various dimensions.

Recently, a class of 1+1 dimensional super
Yang-Mills theories has been studied using
a supersymmetric form of Discrete Light-Cone Quantization
(SDLCQ) \cite{sakai95, alp98, alpII98, alp99}.
This formulation has the  advantage of preserving
supersymmetry after discretizing momenta, and admits
a very natural and straightforward algorithm for
extracting numerical bound state masses and wave functions
\cite{pb85, bpp98}.
Although a technical necessity, the omission
of zero-momentum modes in these numerical computations
raises many doubts about the consistency of such a  
quantization scheme.
Little is in fact known about the precise effects of dropping the 
zero-momentum mode at finite
compactification radius,
but it is generally believed that such effects disappear in the
decompactified limit \cite{alp99}. 

There are instances, however, when we would like
to know the measurable effects of
a finite spatial compactification \cite{dvv97}.
In this work,
we will deal with measurable effects
that reflect the spatial compactification induced
by DLCQ. This is accomplished by explicitly 
incorporating the gauge zero mode
in the DLCQ formulation of a supersymmetric gauge theory.
It turns out that this is tantamount to including a quantum mechanical 
degree of freedom corresponding to 
`quantized electric flux' around the compact direction. 
The implications of this on the vacuum structure of the theory
is discussed. 

The supersymmetric gauge theory we consider may be
obtained by dimensionally reducing ${\cal N} =1$ super-Yang-Mills
from 2+1 to 1+1 dimensions \cite{sakai95}. For simplicity, 
we choose SU(2) to be the gauge group.
The DLCQ formulation of this theory 
consists of an adjoint scalar field (represented as a
$2 \times 2$ Hermitian matrix field), a corresponding
adjoint fermion field, and
several zero-mode (or quantum mechanical) degrees of freedom to be
discussed
later.  To maintain supersymmetry one must impose 
periodic
boundary conditions, so all three color degrees of freedom\footnote{
i.e. arising from the three generators of SU(2).} of the 
fermion and boson fields will have zero modes. 
In addition, periodic boundary conditions prevent us from
adopting the light-cone gauge, $A^+=0$, so we choose
instead the gauge $\partial_-A^+=0$, which allows $A^+$ to have a zero
mode.

For the gauge group SU(2), the field degrees of freedom may be labeled
as
$+$, $-$ and $3$, corresponding to the three generators of the group.
In general, the $\pm$ components of the matter fields
depend on the gauge zero mode and exhibit a spectral flow under large
gauge transformations. In addition, there is a transformation that 
combines a large gauge transformation and a Weyl transformation, and is
known
to be a symmetry of the theory. 
There are two 
important consequences of this. The zero modes
of the $\pm$ components of the fermion and boson fields transform 
into non-zero momentum modes, and 
give rise to a degenerate vacuum in the theory. For this reason it is
inconsistent to omit these modes, and so we will formally 
include them
in our formulation. We will not, however, discuss the 
dynamical implications of including
such degrees of freedom, although some discussion on this and 
related issues appeared
recently \cite{alp99}.

In this work, we concentrate on the effect of including
the quantum mechanical degree of freedom represented
by the gauge zero mode. This zero mode corresponds to
a quantized color electric flux that circulates around the compact
direction
$x^-$.  The problems associated with this zero mode have already been
studied
in two dimensional gauge theories involving
adjoint scalars, and theories with adjoint fermions
\cite{kpp94,pk96,mrp97,kall,pin97a}. The consequences
of including these modes are quite drastic. These theories are known to
possess anomalies in the 3-component of the current,
and a simple consequence is that the charges in these theories 
are time dependent. This makes it difficult -- if not impossible -- to
define a consistent theory. In contrast, owing to special
supersymmetric cancellations between boson and fermion currents,
no such anomalies arise in the supersymmetric theory studied here,
and so a DLCQ formulation becomes sensible and tractable. 

In general, one finds a contribution after normal ordering the 
Hamiltonian that is a function only of 
the gauge zero-mode. This term acts as a vacuum potential and 
leads to a non-zero vacuum energy.  When the gauge theory
without matter fields is solved, however, the only degree of freedom is
the
quantum mechanical gauge zero mode, in which the vacuum potential plays
no
role. The ground state energy is thus zero.
However, this simple picture of the vacuum may be drastically
altered if we consider the addition of matter. 
For the supersymmetric case studied here, we show that 
there is no vacuum potential, and that the ground state has zero energy 
{\em even in the presence of matter fields.} 

This paper is organized as follows. In Section 2, we briefly
describe the DLCQ procedure of the 1+1 dimensional supersymmetric
Yang-Mills theory in the modified light-cone gauge.
In Section 3, the point splitting regularization designed to
preserve symmetry under large gauge transformations
is applied to the current operator. In Section 4,
we discuss the vacuum structure of the theory by deriving
the quantum mechanics of the gauge zero mode.
We conclude in Section 5 with a brief discussion.

%%%%%%%%%%%%%%%%%%%%%%%%%%%%%%%%%%%%%%%%%%%%%%%
\section{Gauge Fixing in DLCQ}
We consider the supersymmetric Yang-Mills theory in
1+1 dimensions \cite{fer65} which is described by
the action
\begin{equation}
S  =  \int d^2 x \hspace{1mm}
 \mbox{tr} \left(-\frac{1}{2} F_{\mu \nu}F^{\mu \nu}
+D_\mu \phi D^\mu \phi +
i \bar{\Psi}\gamma^\mu D_{\mu}\Psi -ig\phi
\bar{\Psi}\gamma_5\Psi \right),
\end{equation}
where $D_\mu=\partial_\mu+ig[A_\mu, \cdot \hspace{1mm} ]$ and
$F_{\mu\nu}=\partial_\mu A_\nu -\partial_\nu A_\mu
+ig [A_\mu, A_\nu]$. All fields are in the adjoint
representation of the gauge group SU(2). A
convenient  representation of the gamma matrices is $\gamma^0=\sigma^2$,
$\gamma^1=i\sigma^1$ and $\gamma^5=\sigma_3$ where $\sigma^a$ are
the Pauli matrices. In this representation the Majorana spinor
is real. We choose Cartan basis,
$(\tau^+, \tau^-, \tau^3)$, defined by $[\tau^+, \tau^-]=\tau^3$ and
$[\tau^3,\tau^{\pm}] =\pm\tau^{\pm}$, with the normalization ${\rm
tr}(\tau^+\tau^-) ={\rm tr}(\tau^3\tau^3)=1/2$. In terms of this basis
the matrix valued fields may be decomposed as follows:
\begin{equation}
A^{\mu}={A^{\mu}}_+\tau^+ +{A^{\mu}}_-\tau^- +{A^{\mu}}_3\tau^3.
\end{equation}
Henceforth the {\em lower} index refers to the gauge group
component. 

We now introduce the light-cone coordinates 
$x^{\pm}=\frac{1}{\sqrt 2}(x^0 \pm x^1)$.
The longitudinal coordinate $x^-$ is compactified
on a finite interval $x^-\in [-L, L]$ \cite{my76, pb85}
and we impose periodic boundary conditions on all fields
to ensure unbroken supersymmetry.

The light-cone gauge $A^+=0$ can not be used in a finite
compactification radius,
but the modified condition $\partial_-A^+=0$ \cite{kpp94}
is consistent with the light-like  compactification.
We can make a
global rotation in color space so that
the zero mode is diagonalized $V(x^+)=v(x^+)\tau^3$ \cite{kpp94}.
The gauge zero mode corresponds to a (quantized) color electric flux
loop
around the compactified space.

The modified light-cone gauge is not a complete gauge
fixing. In fact, the large gauge transformation
$U(x)={\rm exp}(-\frac{i\pi n x^-}{L}\tau^3)$ generates shifts
in the zero mode
\begin{equation}
v(x^+)\rightarrow v(x^+)+\frac{n\pi}{gL},  \hspace{1cm} n\in {\bf {\rm
Z}},
\label{lgt}
\end{equation}
while preserving the gauge condition $\partial_-A^+=0$. To completely
fix
the gauge one therefore fixes $v$ to be in the finite interval
$0<v<\pi/gL$.
It is convenient to introduce
the dimensionless variable $z=gLv/\pi$ as well.
Other intervals give gauge ``copies''
\cite{gri78}  of this domain which is called the fundamental domain.
In addition, we can explicitly see
that the Wilson loop ${\rm cos}(2Lg v)$,
is invariant under (\ref{lgt}).

With this gauge choice the quantization
is straightforward.
The details of this light-cone formulation
may be found in the literature
\cite{pk96, mrp97, sakai95, alpII98, bpp98}. Here we provide  only the
results which are useful for later purposes.
The quantization proceeds in two steps.
First, we must resolve the constraints
to eliminate the redundant degrees of freedom.
There are two constraints in the theory,
\begin{eqnarray}
&&-{D_-}^2 A^- =gJ^+, \\
&&\sqrt{2} i D_- \chi=g[\phi, \psi],
\end{eqnarray}
where $\Psi\equiv (\psi, \chi)^{{\rm T}}$ and
the current operator is
\begin{equation}
J^+(x)=\frac{1}{i}[\phi(x), D_-\phi(x)]-
\frac{1}{\sqrt 2}\{\psi(x), \psi(x)\}.
\label{current}
\end{equation}
The first
equation is the Gauss-law constraint, and in components takes the form
\begin{eqnarray}
&&-{\partial_-}^2 {A^-}_3=g{J^+}_3, \label{glc3}\\
&&-(\partial_- +igv)^2 {A^-}_+ =g{J^+}_+ \label{glc+},\\
&&-(\partial_- -igv)^2 {A^-}_- =g{J^+}_-\label{glc-}.
\end{eqnarray}
We may therefore eliminate any dependence of
$A^-$ and $\chi$ in favor of the physical degrees of
freedom ($\phi$, $\psi$, $v$). However the kernel
of the operator $D_-$ has to be treated separately.
For example, the zero mode of (\ref{glc3}) which is associated with
the kernel of $\partial_-$ provides us with the condition
$\int dx^- {J^+}_3=0$ which must be imposed on the Fock space
to select the physical states in the quantum theory.

%\begin{eqnarray}
%&&{J^+}_3=\frac{1}{i}(\phi_+\pi_- -  \phi_-\pi_+)
%-\frac{1}{\sqrt2} (\psi_+\psi_- -\psi_-\psi_+), \\
%&&{J^+}_+= ({J^+}_-)^{\dagger}=\frac{1}{i}(\phi_3\pi_+ -  \phi_+\pi_3)
%-\frac{1}{\sqrt2} (\psi_3\psi_+ -\psi_+\psi_3).
%\end{eqnarray}

The next step is to derive the commutation relations
for the physical degrees of freedom.
As in the ordinary quantum mechanics, the zero mode
$v$ has a  conjugate momentum  $p=2L\partial_+v$ and
the commutation relation is $[v, p]=i$ \cite{kpp94}.
The off-diagonal components of the scalar field
are complex valued operators with
$\phi_+=(\phi_-)^{\dagger}$. The canonical momentum fields
conjugate to $\phi_-$ and $\phi_+$ are $\pi_+=(\partial_-+igv)\phi_+$
and $\pi_+=(\partial_- -igv)\phi_+$, respectively.
They satisfy the canonical commutation relations \cite{pk96}
\begin{equation}
[\phi_+(x), \pi_-(y)]_{x^+=y^+}=
[\phi_-(x), \pi_+(y)]_{x^+=y^+}=\frac{i}{2}
\delta(x^--y^-).
\end{equation}
On the other hand, the quantization of the
diagonal component $\phi_3$ needs care. As mentioned in
\cite{pk96}, the zero mode of $\phi_3$, the mode independent
of $x^-$, is not an independent degree of freedom but
obeys a certain constrained equation \cite{my76, pk96, kall}.
Except the zero
mode, the commutation relation is canonical
\begin{equation}
[\phi_3(x), \partial_-\phi_3(y)]_{x^+=y^+}=
\frac{i}{2}
\left[\delta(x^--y^-)-\frac{1}{2L}\right].
\end{equation}
Finally, the canonical anti-commutation relations which we
should impose on fermion fields are \cite{mrp97}
\begin{equation}
\{\psi_+(x), \psi_-(y)\}_{x^+=y^+}=
\frac{1}{\sqrt 2}
\delta(x^--y^-).
\end{equation}
We will also omit the zero mode of $\psi_3$, but simply note here that
it has a special role in ensuring supersymmetry. 
This will be discussed in future work \cite{alp99a}.
The corresponding relations are then
\begin{equation}
\{\psi_3(x), \psi_3(y)\}_{x^+=y^+}=
\frac{1}{\sqrt 2}\left[\delta(x^--y^-)-\frac{1}{2L}\right].
\end{equation}

The commutation relations for the Fourier modes, and
the form of 
the light-cone Hamiltonian will be given in the
following Sections.

%%%%%%%%%%%%%%%%%%%%%%%%%%%%%%%%%%%%%%%%%%%%%%%
\section{Current Operators}
The resolution of the Gauss-law constraint
is a necessary step for obtaining
the light-cone Hamiltonian.
The expression for the current operator is,
however, ill-defined unless an appropriate definition
is specified,  since 
the operator products are defined at the same point.
We shall use the point-splitting regularization
which respects the symmetry of the theory under the
large gauge transformation.

The mode-expanded fields at the light-cone time $x^+=0$ are
\begin{eqnarray}
\phi_-(x)&=& (\phi_+(x))^{\dagger} =\frac{1}{\sqrt{4\pi}}
\left( \sum_{n=0}^{\infty} b_n u_n {\rm e}^{-ik_n x^-}
+\sum_{n=1}^{\infty} d^{\dagger}_n v_n
{\rm e}^{ik_n x^-}\right), \nonumber \\
&& \phi_3(x)= \frac{1}{\sqrt{4\pi}}
\sum_{n=1}^{\infty}\frac{1}{\sqrt n}\left(a_n {\rm e}^{-ik_n x^-}
+ a^{\dagger}_n {\rm e}^{ik_n x^-}\right), \nonumber \\
\psi_+(x)&=& (\psi_-(x))^{\dagger}=
\frac{1}{2^{\frac{1}{4}}\sqrt{2L}}
\left( \sum_{n=0}^{\infty}B_n {\rm e}^{-ik_n x^-}
+\sum_{n=1}^{\infty} D^{\dagger}_n {\rm e}^{ik_n x^-}\right),
\nonumber \\
&&\psi_3(x)= \frac{1}{2^{\frac{1}{4}}\sqrt{2L}}
\sum_{n=1}^{\infty}\left(A_n {\rm e}^{-ik_n x^-}
+ A^{\dagger}_n {\rm e}^{ik_n x^-}\right),
\end{eqnarray}
where  $k_n=n\pi/L$, $u_n=1/\sqrt{\vert n+z \vert}$ and
$v_n=1/\sqrt{\vert n-z \vert}$ \footnote{$u_n$ and $v_n$ are 
well-defined in the 
fundamental domain. Similarly, 
$(\partial_-\pm igv)^2$ in the Gauss-law constraint 
have no zero modes in this domain.}.
The (anti)commutation
relations for Fourier modes are found in \cite{pk96, mrp97}
and take the form
\begin{eqnarray}
&&[b_n, b^{\dagger}_m]={\rm sgn}(n+z)\delta_{n, m}, \quad
[d_n, d^{\dagger}_m]={\rm sgn}(n-z)\delta_{n, m}, \nonumber \\
&&\{B_n, B^{\dagger}_m\}=\{D_n, D^{\dagger}_m\}=\delta_{n, m} \quad
[a_n, a^{\dagger}_m]=\{A_n, A^{\dagger}_m\}=\delta_{n, m},
\end{eqnarray}

The large gauge transformations are  
denoted by $T_n$ which act  as  $ T_n z  T_n^{-1}=z+n$.
The charged fields receive the phase rotations
\begin{eqnarray}
&& T_n \phi_{\pm} T_n^{-1}= \phi_{\pm}{\rm e}^{\mp \frac{i\pi n}{L}x^-},
\\
&& T_n \psi_{\pm} T_n^{-1}=\psi_{\pm}{\rm e}^{\mp \frac{i\pi n}{L}x^-},
\end{eqnarray}
while the color neutral fields, $\phi_3$ and $\psi_3$, are unchanged
under $T_n$. One can observe that the large gauge transformation
preserves periodic boundary conditions and
are tantamount
to a spectral
flow. This gauge symmetry is an example of the
Gribov ambiguity \cite{gri78} and can be fixed once we
restrict ourselves to one of the ``copies''.  The fundamental
domain $0<z<1$ is one of such gauge choices,
and this completes the gauge fixing.

The theory has another symmetry called
the Weyl conjugation symmetry, denoted by $R$,
\begin{equation}
R z R^{-1}=-z, \quad R \phi_{\pm} R^{-1}=\phi_{\mp},
\quad  {\rm and}  \quad R \psi_{\pm} R^{-1}= \psi_{\mp}.
\end{equation}
Although the  Weyl symmetry is no longer
the symmetry of the theory after the ``gauge fixing''
%In terms of modes, $R b_n R^{-1}= R^{-1}  d_n$ ($n\ge 1$),
%$R b_0 R^{-1}=b_0^{\dagger}$ for the boson and
%$R  B_n R^{-1}=D_n$ ($n\ge 1$),
%$R B_0 R^{-1}=B_0^{\dagger}$ for the fermion.
there still exists a symmetry of the gauge-fixed
theory which is a particular combination of the large
gauge transformation and the Weyl conjugation
$ S=T_1R $. In fact $S$ maps the fundamental domain onto itself.
This operator can be chosen to satisfy $S^2=1$
and is used in classifying the vacua \cite{lst95, pin97a}.

Let us now discuss the definition of singular operator products
in the current (\ref{current}). We define the current operator
by point splitting,
\begin{equation}
J^+ \equiv {\rm lim}_{\epsilon \rightarrow 0}\left( {J^+}_{\phi}(x;
\epsilon)
+{J^+}_{\psi}(x; \epsilon) \right),
\label{current2}
\end{equation}
where the divided pieces are given by
\begin{eqnarray}
&& {J^+}_{\phi}(x; \epsilon)=\frac{1}{i}
\left[{\rm e}^{-i\frac{\pi z \epsilon}{L}\tau^3}
\phi(x^- -\epsilon){\rm e}^{i\frac{\pi z \epsilon}{L}\tau^3},
D_-\phi(x^-)\right] \label{bcurrent}\\
&& {J^+}_{\psi}(x; \epsilon)=-\frac{1}{\sqrt 2}
\left\{ {\rm e}^{-i\frac{\pi z \epsilon}{L}\tau^3}
\psi(x^--\epsilon){\rm e}^{i\frac{\pi z \epsilon}{L}\tau^3},
\psi(x^-)\right\}.
\label{fcurrent}
\end{eqnarray}
An advantage of this regularization is that
the current transforms covariantly under the
large gauge transformation.  In fact, it is easy to
show that ${J^+}_3$
%\begin{eqnarray}
%{J^+}_3 &=&\frac{1}{i}\left[ \phi_+(x^--\epsilon)
%\pi_-(x^-){\rm e}^{-i\frac{\pi z \epsilon}{L}}
%-\phi_-(x^--\epsilon)\pi_+(x^-)
%{\rm e}^{i\frac{\pi z \epsilon}{L}}\right] \\
%&-&\frac{1}{\sqrt 2}\left[ \psi_+(x^--\epsilon)
%\psi_-(x^-){\rm e}^{-i\frac{\pi z \epsilon}{L}}
%-\psi_-(x^--\epsilon)\psi_+(x^-)
%{\rm e}^{i\frac{\pi z \epsilon}{L}}\right],
%\end{eqnarray}
is invariant under the large gauge transformation
while the others transform covariantly as
${J^+}_{\pm}\rightarrow {J^+}_{\pm}
{\rm e}^{\mp i\frac{\pi n}{L}x^-}$.

It is straightforward to evaluate (\ref{bcurrent})
and (\ref{fcurrent})  and they have previously been discussed separately
\cite{pk96,mrp97}. With a slight modification they are found to be
\begin{eqnarray}
&& {\rm lim}_{\epsilon \rightarrow 0}\hspace {1mm} {J^+}_{\phi}(x;
\epsilon)
=\tilde {J^+}_{\phi}(x) -\frac{1}{2L}(-z+\frac{1}{2})\tau^3,
\label{bcurrent2} \\
&&{\rm lim}_{\epsilon \rightarrow 0} \hspace {1mm} {J^+}_{\psi}(x;
\epsilon)
=\tilde {J^+}_{\psi}(x) -\frac{1}{2L}(z+\frac{1}{2})\tau^3,
\label{fcurrent2}
\end{eqnarray}
where $\tilde {J^+}_{\phi}$ and $\tilde {J^+}_{\psi}$ are
the naive normal ordered currents. To be more precise,
we have omitted the zero modes of the color 3
sectors in which the notorious constrained
zero mode \cite{my76} appears. On the other hand,
the zero modes of the color charged sectors
are explicitly incorporated and  their
effects are found in (\ref{bcurrent2}) and
(\ref{fcurrent2}) as constant terms which
are independent of $z$. As can be seen,
$ {J^+}_{\phi}$ and $ {J^+}_{\psi}$
acquire extra $z$ dependent terms, so called gauge corrections.
Integrating these charges over $x^-$, one finds that the charges are
time
dependent. Of course this is an unacceptable situation, and 
implies the need to impose
special conditions to single out `physical states' to form 
a sensible theory. The
important simplification of the  supersymmetric model is that these time
dependent terms cancel, and the full current  (\ref{current2})
becomes
\begin{equation}
J^+(x)=\tilde {J^+}_{\phi}+ \tilde {J^+}_{\psi}
-\frac{1}{2L}\tau^3.
\label{nocurrent}
\end{equation}
The regularized current is thus equivalent to the naive
normal ordered current up to an irrelevant constant.
Similarly, one can show that $P^+$ picks up gauge correction when the
adjoint scalar or adjoint fermion are considered separately but in the
supersymmetric theory it is nothing more than  the expected normal
ordered
contribution of the matter fields.

In one sense these results are a consequence of the well known fact that
the
normal ordering constants in a supersymmetric theory cancel between 
fermion and boson contributions. The important point here is that these
normal
ordered constants are not actually constants, but rather quantum
mechanical
degrees of freedom. It is therefore not obvious that they should
cancel. Of course, this property profoundly effects the 
dynamics of the theory.
%%%%%%%%%%%%%%%%%%%%%%%%%%%%%%%%%%%%%%%%%%%%%%%
\section{Vacuum Energy}
The wave function of the vacuum state for the
supersymmetric Yang-Mills theory in 1+1 dimensions 
has already been discussed in
the
equal-time formulation \cite{oda95}.
An effective potential is computed in a weak
coupling region as a function of the gauge zero mode
by using the adiabatic  approximation.
Here we analyze the vacuum structure of the same theory
in the context of the DLCQ formulation.

The presence of zero modes renders the light-cone
vacuum quite nontrivial, but the advantage of the light-cone
quantization
becomes  evident: the ground state is the Fock vacuum for a fixed
gauge zero mode and therefore our ground state may be
written in the tensor product form
\begin{equation}
\vert \Omega \rangle \equiv \Phi[z]\otimes
\vert 0 \rangle,
\label{vacuum}
\end{equation}
where we have taken the  Schroedinger representation
for the quantum mechanical degree of freedom $z$
which is defined in the fundamental domain.
In contrast, to find the ground state of the fermion
and boson for a fixed value of the gauge zero mode
turns out to be a highly nontrivial task in the equal-time
formulation \cite{oda95}.

Our next task is to derive an effective Hamiltonian
acting on $\Phi[z]$. The light-cone Hamiltonian
$H \equiv P^-$ is obtained from energy momentum tensors,
or through the canonical procedure.
In terms of the dimensionless operator
$\hat{H}\equiv \frac{4\pi^2}{g^2L}H$,
it is schematically given by
\begin{eqnarray}
\hat{H} &=&-\frac{1}{{\rm sin}^2(\pi z)}\frac{\partial}{\partial z}
{\rm sin}^2(\pi z)\frac{\partial}{\partial z}, \\
&+& \frac{4\pi^2}{g^2L}\int_{-L}^L dx^-{\rm tr} \left(
-g^2 J^+\frac{1}{D_-^2} J^+
+\frac{ig^2}{\sqrt 2}[\phi, \psi]
\frac{1}{D_-}[\phi, \psi]\right),
\end{eqnarray}
where the first term is the kinetic energy of the
gauge zero mode,
and in the second term the zero modes of $D_-$ are understood
to be removed. Note that the kinetic term of the gauge
zero mode is not the standard form $-d^2/dz^2$ but acquires a nontrivial
Jacobian which is nothing but the Haar measure of SU(2).
The Jacobian originates from the unitary transformation of the variable
from
$A^+$ to $v$, and can be derived  by explicit evaluation of a functional
determinant \cite{lnt94, lst95}. In the present context
it is found in \cite{kall}.

Projecting the light-cone
Hamiltonian onto the Fock vacuum sector we obtain
the quantum mechanical Hamiltonian
\begin{equation}
\hat{H}_0=-\frac{1}{{\rm sin}^2(\pi z)}\frac{\partial}{\partial z}
{\rm sin}^2(\pi z)\frac{\partial}{\partial z}
+V_{JJ}+V_{\phi\psi},
\end{equation}
where the reduced potentials are defined by
\begin{eqnarray}
&& V_{JJ}\equiv -\frac{4\pi^2}{L} \int_{-L}^L dx^-
\langle {\rm tr} J^+\frac{1}{D_-^2} J^+  \rangle, \\
&&V_{\phi\psi}\equiv\frac{4i\pi^2}{\sqrt 2L} \int_{-L}^L dx^-
\langle {\rm tr}
[\phi, \psi]
\frac{1}{D_-}[\phi, \psi]\rangle,
\end{eqnarray}
respectively. As stated in the previous section,
the gauge invariantly regularized current
turns out to be precisely the normal ordered current
in the absence of the zero modes.
It is now straightforward to evaluate $V_{JJ}$ and $V_{\phi\psi}$
in terms of modes. One finds that they cancel among themselves
as expected from the supersymmetry:
\begin{equation}
V_{JJ} =-V_{\phi\psi}=\frac{1}{4}\sum_{n, m=1}\left[
\frac{1}{n(m+z)}+\frac{1}{m(n-z)}+\frac{1}{(n-z)(m+z)}\right].
\end{equation}
Thus we arrive at
\begin{equation}
\hat{H}_0=-\frac{1}{ {\rm sin}^2(\pi z)}\frac{\partial}{\partial z}
{\rm sin}^2(\pi z)\frac{\partial}{\partial z}.
\label{h0}
\end{equation}
In order to solve the eigenvalue problem,
it is useful to define the new wavefuntion
\cite{lnt94, lst95, oda95}
\begin{equation}
\tilde{\Phi}[z]={\rm sin} (\pi z)\Phi[z],
\end{equation}
in analogy with the radial wavefunctions.
The appearance of the Jacobian seems
vital since it determines the behavior of the $\tilde{\Phi}[z]$
at the edges of the fundamental domain: it leads to the
boundary condition \cite{lnt94, lst95}
\begin{equation}
\tilde{\Phi}[z=0]= \tilde{\Phi}[z=1]=0.
\end{equation}
The eigenvalue problem $\hat{H}_0\Phi[z]=E\Phi[z]$ now turns out to be
\begin{equation}
\left(-\frac{d^2}{dz^2} -\pi^2 \right)
\tilde{\Phi}[z]=E  \tilde{\Phi}[z],
\label{eigen}
\end{equation}
which can be solved easily. One can find that the ground state energy
is precisely zero $E_0=0$ and the corresponding vacuum wave
function is
\begin{equation}
\tilde{\Phi}[z]=\sqrt 2{\rm sin}(\pi z).
\end{equation}
We have thus found within the present assumption
that the ground state has a vanishing vacuum energy, suggesting that
the supersymmetry is not broken spontaneously.

Note that an emergence of the ``effective potential''
$-\pi^2$ in (\ref{eigen}) is essential to  this conclusion.
In non-supersymmetric theories this constant
energy is simply disregarded since it merely shifts
all energy eigenvalues by the same amount. Then the
eigenvalue problem (\ref{eigen}) becomes formally the same as the
original
one $\hat{H}_0\Phi[z]=E\Phi[z]$ but with the standard kinetic
term $-d^2/dz^2$ supplemented by the boundary condition
$\Phi[z=0]=\Phi[z=1]=0$ (not for $\tilde{\Phi}[z]$).
In supersymmetric theories, however, such constant
energy cannot be discarded by hand since it is a part
of dynamics closely related to the vacuum structure.

\section{Discussion}
We have performed a detailed analysis
of the quantum mechanical degrees of freedom represented
by the gauge zero mode of a supersymmetric SU(2) gauge theory 
in $1+1$ dimensions. This theory may be obtained by dimensionally
reducing ${\cal N}=1$ super-Yang-Mills from $2+1$ dimensions, and
consists
of one adjoint fermion and one adjoint scalar field periodically
identified in the $x^-$ light-cone coordinate. 

The remaining zero mode degrees of freedom
that were not treated here are known to give rise
to vacuum effects that are crucial in understanding 
the effects of topological field configurations
due to the non-trivial center of SU(2),  and the
existence of two-fold vacua
$\vert \Omega_\pm \rangle$,  along
the lines investigated in \cite{pin97a}. 
Evidently, an understanding of all the zero-mode degrees of
freedom awaits future work.

We have, however, carefully taken
account of the gauge zero mode.  In this context the gauge zero mode is
a
quantum mechanical degree of freedom. In general, when one
normal orders the operators of the theory one finds contributions that
depend
only on this quantum mechanical degree of freedom.  These term are
anomalies
and profoundly effect the structure of the theory. In theories with only
fermions or only bosons, these anomalies yield time
dependent charges and a non-zero vacuum energy. In the supersymmetric
theory
presented here, these anomalies are seen to cancel and the operators 
are all well behaved. In particular,
the charges are time independent and the ground state with matter 
is the same state 
that one finds when the theory is quantized without matter. In as much
as the
ground state energy is zero we conclude that the gauge zero mode does
not break
supersymmetry.

Finally, we remark that the properties of Matrix String Theory
\cite{dvv97} -- which is defined as 1+1 ${\cal N}=8$ super-Yang-Mills
theory on a circle -- depend crucially on the measurable effects
produced
by the space-like compactification. These effects are intimately
tied with the dynamics of non-perturbative objects in Type IIA string
theory
known as D0 branes. It would be interesting to consider the
DLCQ formulation of the same Yang-Mills theory, and to establish -- 
if possible -- any connection with the Matrix String proposal.
The simplicity of the light-cone Fock vacuum, owing to
special supersymmetry cancellations, might present a tractable approach
to non-perturbative string theory.

%%%%%%%%%%%%%%%%%%%%%%%%%%%%%%%%%%%%%%%%%%%%%%%%%%%%%%

\medskip
\noindent
\begin{large}
{\bf Acknowledgments}\\
\end{large}
S.T. wishes to thank S.Tanimura for helpful discussions.
S.T. would like to thank Ohio State University for hospitality,
where this work was begun.
%%%%%%%%%%%%%%%%%%%%%%%%%%%%%%%%%%%%%%%%%%%%%%%%%%%%%%%%%%
%%%%%%%%%%%%%%%%%%%%%%%%%%%%%%%%%%%%%%%%%%%%%%%%%%%%%%%%%%
%\begin{eqnarray}
%A(z)&=&\frac{g^2L}{4\pi^2}\sum_{n, m=1}\left[ \frac{}{} \right.
%\frac{1}{(n+m+z)^2}+\frac{1}{(n+m-z)^2}+\frac{1}{(n+m)^2}
%\left. \frac{}{} \right]
%\nonumber \\
%&+& \frac{g^2L}{16\pi^2}\sum_{n, m=1}\left[ \frac{}{} \right.
%\frac{(n-m+z)^2 }{n(m-z)(n+m-z)^2}+
%\frac{(n-m-z)^2 }{n(m+z)(n+m+z)^2} \nonumber \\
%&+&
%\frac{(n-m+2z)^2 }{(n+z)(m-z)(n+m)^2} \left. \frac{}{} \right],\\
%&=&\frac{g^2L}{16\pi^2}\sum_{n, m=1}\left[
%\frac{1}{n(m+z)}+\frac{1}{m(n-z)}+\frac{1}{(n-z)(m+z)}\right],  \\
%&=&-B(z).
%\end{eqnarray}
%\newpage

%%%%%%%%%%%%%%%%%%%%%%%%%%%%%%%%%%%%%%%%%%%%%%%%%%%%%%%

%

\vfil

\end{document}